# IBM Research Report

## Modeling IoT-aware Business Processes
### A State of the Art Report

Nadja Brouns[1], Samir Tata[2], Heiko Ludwig[2], E. Serral Asensio[1, 3], Paul Grefen[1]

[1]Eindhoven University of Technology
5612 AZ Eindhoven
Netherlands

[2]IBM Research Division
Almaden Research Center
650 Harry Road
San Jose, CA 95120-6099
USA

[3]KU Leuven
Naamsestraat 69
Leuven, Belgium

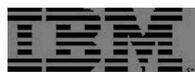

# Modeling IoT-aware Business Processes

## A State of the Art Report

N. Brouns, S. Tata, H. Ludwig, E. Serral Asensio, P. Grefen

The Internet of Things (IoT) is an emerging technology. Many entities see opportunities to use IoT and develop their business. The IoT smart devices will change the everyday life of people. Research mainly has been focusing on technical implementation of IoT and not of the integration into business processes. As there is a clear demand of businesses to have a seamless integration, research has been conducted recently on this topic. This document describes the current state of research in the field of IoT-aware business process modelling.

# Contents

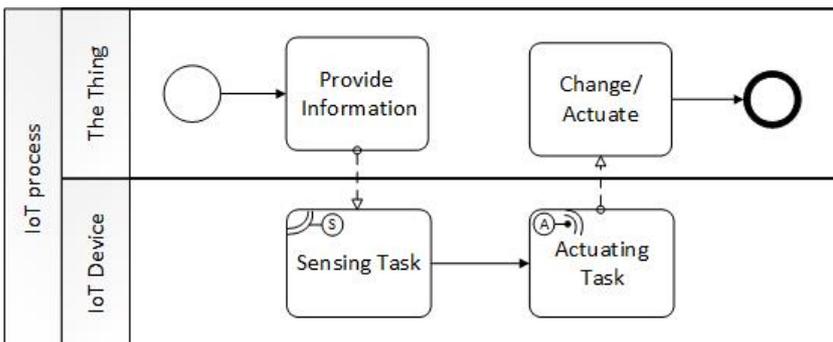

# List of Figures

# List of Tables

# 1 Introduction

This report presents an analysis of the state of the art of modeling IoT-aware business processes. In this introductory section, we first explain why it is relevant to combine the Internet of Things (IoT) and business process management (BPM). This sets the stage for the rest of this report. Next, we explain the goals and the structure of this report.

## 1.1 The relevance of combining IoT and BPM

The Internet of Things is a development that links the physical world to the digital world. Traditionally, we would find information about events and processes in the physical world in the digital world. However, this information is traditionally based on data entered into information systems by humans and humans use this information to control the physical world. In the IoT paradigm, the physical world is equipped with sensors and actuators to create a direct link with the digital world. Sensors allow for direct recording of data into digital systems - at the moment and location of creation of this data. Actuators allow digital systems to exert direct control over the physical world - without the need to have humans as gateways in between.

The Internet of Things (IoT) is often seen as a disruptive technology [1]. Using smart devices it has the possibility to change everyone's daily life. Through large sets of advanced sensors and actuators, it can create opportunities for business organizations to establish new business models. It is not surprising that in the past years the amount of research conducted on the topic of IoT has steadily increased. Until now, most research has focused on technical implications and challenges of IoT. Much research focuses on the design of IoT infrastructures, paying attention to basic communication technology, interoperability of technologies and processing of large amounts of data generated by IoT devices. Other research efforts pay attention to important non-functional aspects, such as security, privacy and scalability of IoT infrastructures.

To deploy IoT concepts in business domains, however, the basic functionality of IoT has to be integrated with the way business organizations work. In other words, activities that are based on IoT have to be integrated into business processes [1]. Business processes define how organizations achieve their goals, and the 'bridge' between these processes and IoT concepts defines how events in the physical world play a role in this [36]. Once we model this bridge, we obtain IoT-aware business processes. An essential observation here is that typical IoT-aware business processes are not linked to a single IoT 'thing', but may require access to many 'things'. The role of the business processes is then the synchronization of these 'things' towards the business goal they serve. Business process modelling is considered a successful solution for coordinated management and optimized resource planning [2]. Bringing it to the IoT world makes the management and planning of physical resources more direct.

IoT can be applied into different areas such as healthcare, transportation, and manufacturing. Although each application domain poses different modeling requirements, it is important to create a uniform way of representing IoT in business process models [3]. Current standardized business process modelling languages cannot represent the real-world aspects that come along with IoT. Languages like BPMN have as goal "to create a standardized bridge for the gap between the business process design and process implementation" [4]. Where these modelling languages focus on process steps and flows in the digital world, IoT focuses on the connection between digital systems and physical agents. Thus, there is a problem when the concepts of IoT and BPM need to directly mapped. The worlds of business process modelling and the Internet of Things seem to be disconnected, while the industry is demanding for a seamless integration of the real world into business processes.

## 1.2 Goal and structure of this report

In the past few years, we have seen a collection of research efforts on the possibilities to model IoT-aware business processes. This collection is relatively small when compared to the overall research effort into the IoT and much of the work is still in the early research stage. To create a basis for the mentioned bridge between IoT and BPM, the goal of this report is to investigate these efforts and to analyze the state of the art of existing frameworks for modeling IoT-aware business processes.

This report is structured as follows. In the next section, we start with exploring possible business domains for the combined application of the Internet of Things and business process management. Section 3 describes the requirements for modeling IoT-aware business processes. Based on these requirements, we analyze several popular business process modeling languages in Section 4 to identify which language can best be adopted as a basis for modeling IoT-aware business processes. We arrive at the conclusion that BPMN 2.0 is the most suitable language, and consequently take this as the basis for the rest of this report. Existing Business Process Modelling Notation (BPMN) elements are described and used for explanation. A basic knowledge of BPMN is required from the reader as these standard BPMN elements are not elaborated on. However, in Appendix A an overview of the main BPMN elements can be found.

In Sections 5 to 10, we investigate the state of the art of modeling business process aspects that are crucial for an IoT context - and that may have different requirements when compared to 'traditional' business processes. We show how these can be modeled as extensions to BPMN 2.0. Section 5 discusses control flow and activity primitives. Section 6 discusses agents that execute business processes - where the 'things' of IoT obviously play an important role. Section 7 treats events, as these are important for the real-time, event-driven nature of IoT applications. Section 8 discusses the role of data in IoT-aware business processes. Sections 9 and 10 finally treat two aspects that often get little attention in traditional business processes, but are essential for IoT applications: location and timing specification.

In Section 11, we present an overview of related work. We conclude this report in Section 12.

# 2 Application Domains

Many business domains can benefit from the implementation of the combination of the Internet of Things and business process management. In this section, we discuss a few major application domains in this section – without any attempt to be complete.

## 2.1 Logistics and freight transport

The development of the IoT started with the use of RFID technology, which has been widely adopted in logistics and retail. Logistics was one of the focal points when IoT started to develop, because it was expected to be very beneficial in this domain. One of the applications is the monitoring of containers in a supply chain [7]. Most container transportation uses multiple modalities and therefore containers have to change transportation modes, which takes time. Time and money can be saved when the container is tracked in a real-time fashion. By tracking the container, its exact location is known and an accurate forecast can be made of its estimated time of arrival at the next location. With this knowledge, one can ensure that the next mode of transportation is ready to leave as soon as the container is placed on the vehicle, so the minimum amount of time is lost.

Furthermore, due to smart information systems and the increased sensing and processing capabilities of vehicles, such as a truck, IoT can be used to identify smart routes. The smart routes can be routes where there is less traffic or where there is a parking space available when the truck driver needs to rest. This can all lead to less congestion on the road and therefore to a more efficient transportation of goods [8].

In current practice, many IoT-based solutions in logistics and transport address individual activities. By integrating these into end-to-end business processes, real-time synchronization along the entire transport chain can be realized and logistics efficiency can be improved.

## 2.2 Healthcare

There is a growing desire along elderly to live independent for a longer time. Internet of Things and Artificial Intelligence can help realize a scenario where people grow old in their own home. Ambient Assisted Living (AAL) provides the opportunity to incorporate these technological inventions [5]. A possible use case is the following.

An elderly woman is monitored from a remote distance using multiple sensors such as GPS, heartrate and blood pressure. The moment she faints the sensors will notice a change and when a reading drops below certain levels or there is no more movement, an alarm is triggered. This alarm notifies a healthcare professional who is monitoring her from the hospital and he sends a notification to the ambulance service. The ambulance receives the information automatically and uses sensors spread over the city to find the fastest route. When they arrive, they are able to help the patient, as the doors of her house are automatically unlocked after the alarm was triggered. From this example, it becomes clear that a number of actions have to be managed in their execution in the form of a business process.

IoT can also be used inside hospitals to track scarce equipment in certain departments of the hospital. Using the data that comes available from tracking the equipment it is also possible to identify if additional equipment is needed due to a high scarcity [6].

## 2.3   Manufacturing

The use of smart robots in factories is getting more common. In the manufacturing environment, these robots often work together with humans. However due to safety restrictions this collaboration is far from efficient. An example research effort addressing these issues is the European HORSE project. The HORSE system brings the worlds of robotics, i.e. the physical world, and the digital world of business information systems together [37]. In this approach, a robot has sensors that provide information to the digital world, where the data is processed. This can lead to a trigger or action in the real world [9], like actuating the robot. An example can be that when the robot is active it senses that a human is standing too close to its moving arm and can get hurt. The information that is fed to the digital world is processed and an actuation task is sent. The robot stops moving its arm, to prevent the human employee from getting hurt.

In traditional IoT applications in manufacturing, the scope is limited to single manufacturing work cells, in which smart machines like robots perform their work. By integrating the activities in multiple work cells into a business process, real-time management of efficiency and effectiveness becomes possible in an end-to-end fashion.

## 2.4   Mobility

The Internet of Things can have a big impact on our everyday transportation. In the past years an increasing amount of research is conducted on cooperative intelligent transportation systems (C-ITS). C-ITS is based on real-time sharing of information between the users of transport infrastructures and traffic managers. The cooperation will help the users to take the right decisions on the right moment.[1] Examples are car drivers that produce real-time data with their cars (that thus become IoT 'things') and receive traffic information on an on-board unit, and travellers in public transportation systems that share their travel plans and whereabouts and receive real-time, individualized travel advice across multiple transportation modes. In making information and advice pertinent to a complete travel process in a real-time way, the interaction between users and multiple infrastructure managers needs to be modeled in a business process.

The C-Mobile project is an example research effort that focuses on deploying C-ITS on a large scale across Europe [38], demonstrating the functionalities and economic viability. Aligning the objectives of the multiple C-ITS will help to improve road transportation, by making it safer, more effective and sustainable.[2]

---

[1] *https://ec.europa.eu/transport/themes/its/c-its_en*
[2] *http://c-mobile-project.eu/*

# 3 IoT Requirements Analysis

This section discusses the requirements for an IoT-aware business process as identified by Meyer et al. in [10]. These requirements are divided into three different categories: 1) the general notation aspects, 2) the aspects for the process editor and 3) the IoT specific aspects. The requirements for the general process notation are setup due to the large variety of possible business process modelling languages. Not all languages might be suitable for modeling IoT-aware business processes and therefore should be evaluated according to the requirements identified by Meyer et al. [10]. However, the focus of this report lies on the evaluation of the requirements from the third category; the IoT specific aspects. Therefore, the requirements for both the first and second category are not listed here.

The requirements that are relevant for this report are the IoT specific aspects. Business processes in an IoT environment must be able to respond quickly to changing requirements and variables, including the integration of real-world technologies. It is important that the IoT specific properties of smart devices such as sensors and actuators are are available in the BPM language and can be represented graphically in the business processes model. The following list defines several requirements for IoT specific process properties, over and above typical BPM.

- The extension must support entity-based modeling and include participants such as a physical entity and an IoT device.
- The extension must support multiple devices to participate in the process, as several devices can execute the different steps in a business process;
- The extension must enable modelling different IoT interactions, such as interactions between devices, services and the use of data;
- The extension must make data storage available and accessible on different devices;
- The extension must allow for a scalable model, such that additional services can be added and their impact is measurable;
- The extension should allow for the abstraction of individual process components. The accumulated data of multiple resources can be more important than the individual data and therefore the individual process components should be abstractive;
- The extension must allow individual components to express its availability, as devices can be mobile the process can depend on the availability of the components;
- The extension should express a tolerable error rate, when this error rate is not yet reach the process can still be modelled;
- The extension must enable a context-aware BPs, such that the process can respond flexible on certain events that occur;
- The extension should express the uncertainty of process components that can influence the creation time of the model;
- The extension should be able to express real-time constraints. The process interacts with the real world, the process should be able to respond and apply these constraints.

The eleven requirements mentioned above are used to analyze the business process modelling languages in the next section. An overview of the analysis is presented in **Table 1**. In some cases, the IoT aspect is already possible in one of the standard notations and no additional graphical elements are necessary to fulfill the requirement. In case the requirement cannot be fulfilled using the standard an extension is necessary. The extensions are mentioned in this report. However, it is not said that the extensions mentioned here will fulfill all requirements. There might still be some research gaps and necessities to model a complete IoT-aware business process model.

# 4 Business Process Modelling Language

This section presents the basic concepts of business process modelling languages and the necessary extension for an IoT-aware notation based on the requirements summarized in the previous chapter. A short definition is presented for the main concepts. Furthermore, a short overview is given of the elements that an overall business process contains.

## 4.1 Standard Business Process Modelling Language

A business process model (BPM) is the representation of a business process. While general graphical process notation such as GANTT Charts reach back to the late 1800s [39], many different executable modeling notations were created when automated business process management systems found adoption in practice more widely, since around the 90s. This enabled companies to implement and deploy modeled processes relatively quickly. A BPM typically also provides a graphical notation. Weske defines a business process as follows: „A business process consists of a set of activities that are performed in coordination in an organizational and technical environment. These activities jointly realize a business goal. Each business process is enacted by a single organization, but it may interact with business processes performed by other organizations" [11].

There are many languages to model business processes. Each of these modelling languages looks very different, but they all consist of three components, syntax, semantics and notation. The syntax ensures that elements are combined correct, by providing elements and a set of rules. The meaning of these elements is provided by the semantics. Lastly, the notation visualizes the language, by providing graphical symbols [12].

All though, the modelling languages contain these three components they are still very different. With the wide variety of business process languages, it is difficult to understand each business process. However, there is a de jure standard released by the Object Management Group (OMG), which is BPMN 2.0. The BPMN2.0 standard is widely used. However, it consists of more than 100 elements, so to fully understand the language it takes time. BPMN contains activities and decision points that are connected, to represent a certain order, by arcs. It is possible to model cross-organizational flowcharts with BPMN, which means that multiple participants are part of the process, using the swimlane notation [12].

Other well-known and widely used modelling languages are the Unified Modeling Language (UML) and Event-drive Process Chains (EPCs). Both languages allow for cross-organizational flowcharts, just like BPMN. However, they both have a very different incentive. A UML Activity Diagram, is mostly used when data plays an important role in the process. In the case a process focuses on events, EPCs is the best choice as events are the core of the modelling language. Every modelling language mentioned above provides the capability to visualize business processes. However, the last modeling language discussed here, does not provide a visualization of a business process model. The Business Process Execution Language (BPEL), is a well-known and often used language and belongs in the same list as the languages mentioned above. However, BPEL does not contain a graphical notation. It is often used for process execution and enables the model that is created in a different language to actually run and be executed, often using a workflow engine [12].

All these modelling languages have different characteristics and can therefore provide a different benefit for IoT-aware business process. In the next section, the Business Process Modelling Languages are compared using the above-defined requirements, in order to see which language fits best for an IoT-aware business process. BPEL is excluded from the comparison, because the focus of this report lies on the visualization of IoT-aware business process models.

## 4.2 Business Process Modeling Language extensions for IoT

Business process modelling can provide great value to the Internet of Things and this also works the other way around. BPM can bring a process perspective to IoT, by putting the IoT data in the right context. The infrastructure provided by BPM can improve the quality and prediction ability of IoT. The other way around, IoT data provides value for both process management and process mining. For example, by discovering unknown processes or processes that happen occasionally [13]. To enhance the full potential of IoT and BPM it is important that IoT is integrated in process models, so IoT-aware processes can be created.

Research suggests that for the upcoming years extending the existing business process modelling languages with IoT-aware elements will be sufficient and therefore, an entirely new BPML is unnecessary yet[10]. This means that the current tendency is to add new objects that represent IoT elements to modelling languages like BPMN and UML. Such an object, can represent multiple things such as a physical thing, or a sensing or actuating task.

Table 1: IoT Characteristics by BPM notations [10]

| IoT characteristics | BPMN 2.0 | eEPC | UML 2.3 |
|---|---|---|---|
| *Entity-based concept* | Partly | No | Partly |
| *Distributed execution* | Partly | No | Partly |
| *Interactions* | No | No | No |
| *Distributed data* | Partly | No | Partly |
| *Scalability* | Partly | No | Partly |
| *Abstraction* | Yes | No | Partly |
| *Availability/Mobility* | No | No | No |
| *Fault tolerance* | Partly | No | Partly |
| *Flexibility/ Event based* | Yes | Partly | Partly |
| *Uncertainty of Information* | No | No | No |
| *Real-time* | Yes | No | No |
| ***Total score*** | **5.5** | **0.5** | **3.5** |

The IoT-A project reviewed three standard notations to identify which is the most suitable for IoT-aware process modelling. BPMN 2.0, eEPC and UML2.3 were evaluated on the eleven requirements discussed in Chapter 3. The results of the evaluation is shown in **Table 1** [10]. The total score is calculated as follows: yes results in one point, partly a half point, and no zero points. The results show that BPMN 2.0 is the most suitable approach to model IoT-aware processes. Furthermore, BPMN 2.0 defines both a graphical and executable notation. An executable notation means that the notation is machine-readable and is mostly presented in an XML format. The XML format is entered into the process engine, which can read the technical aspects and execute the process. Besides, BPMN 2.0 is an extensible modelling language in which new IoT artefacts can be included[10][14]. Lastly, research shows that BPMN language can be used to model end-to-end business processes, which is necessary to use the full potential of the Internet of Things. One of the end-to-end process modelling types is Collaborative B2B Processes. Such a process does not take the view of one single participant, but show the collaboration between participants. In an IoT-aware business process, there will always be interaction between devices and physical entities. An example of a collaborative end-to-end process is presented below [15].

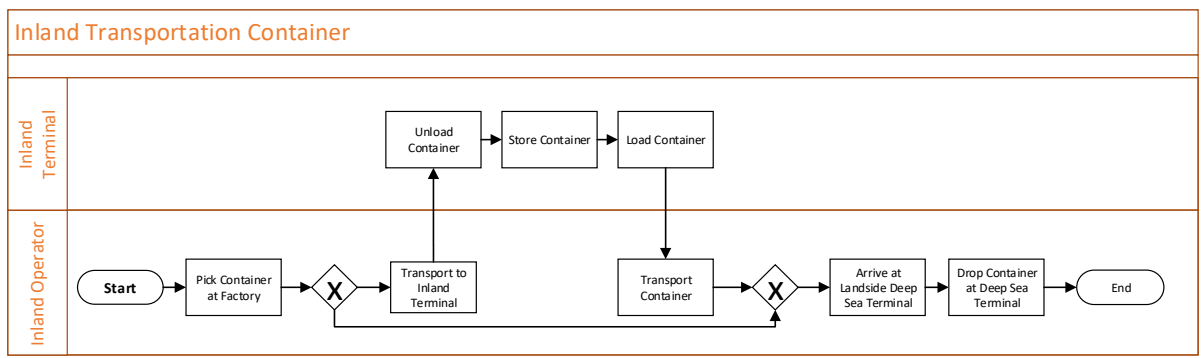

**Figure 1 Collaborative end-to-end business process**

The mapping of important IoT concepts, retrieved from the IoT-A project [10], on the current capabilities of BPMN 2.0 is shown in **Table 2**. BPMN 2.0 does not cover most of the important IoT concepts. An extension for these concepts is necessary in order to model IoT aware processes. Furthermore concepts such as, availability, mobility and information, which were mentioned as requirements for an IoT-aware BPs, are not covered by BPMN 2.0 [10].

**Table 2 IoT Concepts covered by BPMN 2.0 [10]**

| IoT Concept | Covered by BPMN 2.0 |
|---|---|
| (Human) User | ✓ |
| Physical Entity | - |
| Virtual Entity | - |
| Device | - |
| Sensor | - |
| Tag | - |
| Actuator | - |
| Service | ✓ |
| Mobility | - |
| Availability | - |
| Information | - |

The remainder of the report focuses on the extensions for BPMN 2.0 to support IoT aware BPs. These extensions are based on the work of the IoT-A project, but also on other papers from researchers focused on the subject of IoT-aware business processes. The need for these specific extensions is based on the requirements from Section 2 as well as the IoT concepts defined in **Table 2**. Furthermore, the use-cases from the application domain also show the need for the availability in BPs of real-time and secure data, colocation and synchronization of things. Section 5 till 10 present the current available graphical extensions in research for IoT-aware BPs.

# 5 Control Flows for IoT

According to the IOT-A project, we call a process that takes place in the real world among "things" an IOT process. The representation of a business process encompassing an IOT pocess is called "IoT-aware business process". The IoT process consists of two process participants. The first participant is the Physical Entity, also called the thing and is a core element of an IoT-aware business process. The second participant is the IoT Device, which actually has the capability to connect with the Physical Entity, so that it becomes a part of the process. Further elaboration of the process participants is presented in Section 6.

As mentioned above the IoT Device ensures that the Physical Entity can connect with the digital world of the business process. To ensure this connection we need two additional flow elements, the sensing and actuating task. Both elements are needed to define an IoT-aware business process and are discussed in more detail below.

## 5.1 Sensing Task

Sensing is an IoT Device specific task that is usually executed by a sensor. This can be a temperature sensor, heart rate sensor and many more. The sensing task extracts data from the Physical Entity and therefore should have a direct connection to it.

BPMN already offers different types of activities such as, human activities, service activities and receiving activities. However, they do not have the exact same properties as the Sensing activity. Sensing activities have no human intervention, nor are they connected to an external participant, in other words, there are no message flows, and it has no Input Data set. Therefore, additions must be made to BPMN 2.0 in order to visualize the Sensing activity [10]. The graphical notation of the Sensing Activity is given in Figure 2.

SensingActivity is added to the Metamodel and linked to a set of requirements in order to express the necessary properties. Furthermore, it is important that a Sensing Task can be distinguished from the other activities and therefore has its own element.[10] According to Chen & Wang [3] the BPMN should be extended to include which sensor is used to extract data. They create a ResourceRole, which has certain parameters to define which sensor is used. This is only a machine-readable extension, which means that there is no graphical extension to BPMN[3].

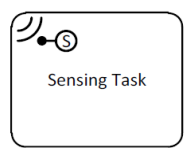

**Figure 2 Sensing Activity**

## 5.2 Actuating Task

The actuator works in the exact reverse direction of a sensor. Instead of observing the environment, it executes a physical action. Most of the time data from the sensor is used to send a command to the actuator to trigger the action. The execution of an actuating task can change the state of the Physical Entity. A simple example would be the unlocking of a door. The state of the door, the Physical entity, changes from locked to unlocked.

BPMN offers the same activities for the Actuation Activity as for the Sensing activity as they have similar properties. In addition, in this case the elements do not completely fulfill the requirements of the Actuation Activity. The element that represents the actuator should be able to adjust the business process. Furthermore a specific relation must be created between the Physical Entity and the Actuation task, namely "acts on"[10].

For the Actuation Activity the same alterations are made to the Metamodel as for the Sensing Activity. There is an input data set that can trigger the action of the actuator and the output data set that verifies if the action was completed successfully. Furthermore, the Actuation Task has an object implementation that specifies to execute a physical task. To distinguish the Actuation Activity from the original activities a graphical element is created, shown in Figure 3[10].

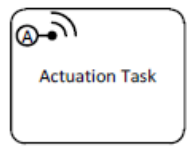

**Figure 3 Actuating Activity**

# 6  IoT Process Participants

This part of the report presents the two process participants that are mandatory for an IoT-aware business process model. As mentioned in the previous section, the definition and representation that is presented here is derived from the IoT-A project, which enjoys wide adoption. Other representations might be viable as well.

The first participant is the physical entity, also often called 'the thing'. Things are the core elements of IoT. These physical objects must be part of the business process in order to call it an IoT-aware business process. The second participant is the IoT Device. The IoT Device is the connection between the Physical Entity and the digital world. Therefore, it is a core element of an IoT-aware business process. Both process participants are discussed in more detail below. In the last part of this section the core elements of the IoT process are visualized, to show the connection between the various elements.

## 6.1  Physical Entity

The Internet of Things is a network where multiple things or objects communicate with each other. No human interaction is needed for this, the network uses the internet and sensors to connect the things[7]. These things are of interest to the business process and when sensors are connected to the thing it becomes part of the digital world[14]. In the IoT-A research the things are defined as a Physical Entity [10].

Business processes mostly consider multiple departments or participants, which are represented by pools or lanes. A Physical Entity can be important for multiple activities spread through the different departments. This can also involve multiple IoT devices within an IoT-aware business process. The Physical Entity should be represented in the business process, so that it can connect to the multiple departments and IoT devices [10].

The current BPMN 2.0 provides Text Annotation, Data Object and Participant as modelling concepts that could represent the Physical Entity. The element Participant represents the Physical Entity best. However, a Physical Entity needs to be connected to the IoT devices, as it does not contain flows and neither can actuate anything. Furthermore, the connection between the Physical Entity and IoT device must be expressed in the business process, but should be distinguished from the sequence flow and message flow. The concept Participant does not fulfill these requirements and therefore a new element is introduced [10][14].

Meyer et al. present both a graphical and a machine-readable element for the new concept Physical Entity [14]. The Physical Entity is represented as an empty lane with horizontal lettering, as it cannot contain any flow elements. The connection to the IoT Device is made, using an information flow. This flow shows the direction in which the information is sent. The new element does not contradict with the restrictions of the standard BPMN 2.0 elements.

Chen & Wang found that the meta model from the IoT-A project is too complicated [3]. BPMN users would take too long to get familiar with these new elements. Therefore, they decided not to explicitly define the Physical Entity. According to Chen & Wang[3] the Physical Entity can be included in to the meta model as a lane or a participant with whom the system interacts.

In the case that the Physical Entity is seen as a thing that also can have an influence on the process, by actuation, movements or just simple changing the state that it is in, it is better to model it as a process participant. When representing it as a lane, which is suggested by Chen & Wang[3], the Physical Thing can have activities of its own and therefore can influence the process. Another way to look at it is the approach of Meyer et al.[14]. They treat the Physical Thing as a black box and an element that is independent from the process. It is questionable what the best approach to model a Physical Entity is. In order to determine the best solution, both approaches should be used in use-cases and must be compared to determine the advantages and disadvantages.

## 6.2 IoT Device

The IoT Device is a substantial part of the IoT-aware business process. It ensures the connection between the physical world and the digital world. The IoT device can execute tasks such as sensing, actuating and monitoring. The artificial link "attached to" between the IoT device and the Physical Entity ensures the latter to be part of the digital world[16]. An IoT device can be anything with a sensor or actuator attached to it, but also for example a mobile phone.

According to Meyer et al. [10] the IoT device performs like a technical process participant. In the current BPMN, a technical process participant is modelled as a pool or a lane. However, this would not fulfill the requirements for representing the IoT device in a business process. It is important that the IoT device can connect to the Physical Entity. Furthermore, the lane of an IoT device should be distinguished from an organizational or business lane.

To integrate the IoT Device into BPMN the subclass IoTDeviceRole is introduced to the metamodel. Besides that, the IoTDeviceRoleParameter and expression are added. This is done to ensure that certain properties can be linked to the IoT Device. Until now, no annotations or requirements can be attached to an individual participant. A lane with a plus symbol graphically represents the IoT Device, so it can collapse and show its parameters. The set of parameters that is linked to that particular IoT device can be used to find the IoT device that fits the process best. Furthermore it distinguished the lane of the IoT Device from the regular participants[10].

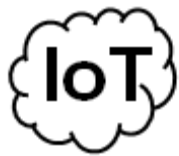

**Figure 4 IoT Device**

**In order to make it easier to distinguish the IoT device from the regular participants, Petrasch & Hentschke [17] introduced a new element. The element shown in Figure 4 is added to the lane that represents the IoT Device. As mentioned before the IoT device can be a mobile phone, which receives a message. In that case, a human can be involved to read that message and act upon it. This could be the case with a patient emergency and the doctor is notified by an alarm on his mobile phone. In order to represent the human involvement another element is created, shown in**

Figure 5 [17].

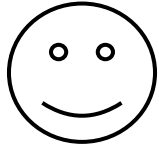

**Figure 5 Human Involvement**

The role of an IoT Device is not only that of a participant but it is a performer as well. An IoT Device can execute multiple activities and multiple activities can use the IoT device. Therefore, the IoT device can also be seen as a resource. In BPMN, a resource classifies the operational capacity during the process. Currently the BPMN applies a large set of parameters to the resource, describing its properties. Parameters can be defined for a participant in order to describe its resource properties. However, we typically cannot to assign resource properties on the activity-level. The IoT device is seen as separate from process participants. We cannot assign parameters to the IoT Device. Hence, the IoT Device cannot be properly treated as a resource even though it is identified as such. To enable resource property allocation on an activity level, there is no graphical element added, only a machine-readable. Furthermore, the meta model is extended by introducing the class IoTPerformer. This is associated to the IoTDeviceRole, which in his place is linked to a resource. This makes it possible to link resources to a specific activity [10].

## 6.3 IoT-aware Business Process

Figure 6 shows an IoT process including one participant, an IoT device, and the thing, as defined by the IoT-A project. The process further includes one sensing task and actuation task and the element for IoT device [10]. In Figure 7 a process model is shown according to the approach of Chen & Wang [3]. Where the thing is modeled as a lane instead of a collapsed pool.

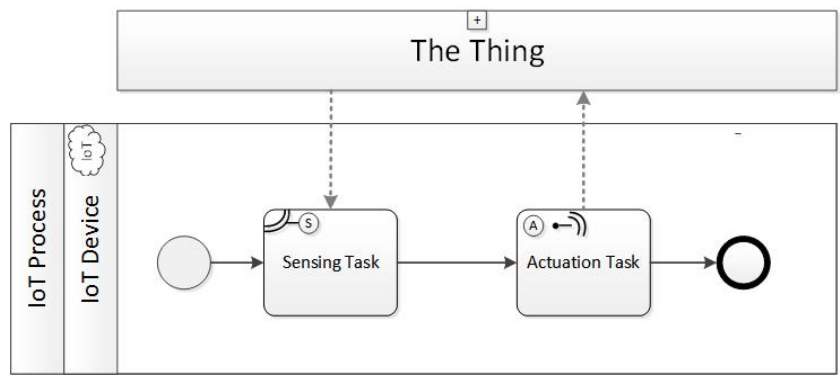

**Figure 6 IoT-aware Business Process according to IoT-A project**

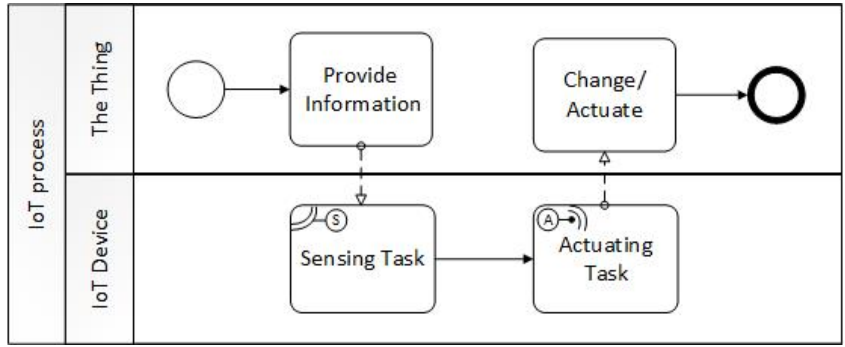

**Figure 7 IoT-aware Business Process with Physical Entity modeled as lane**

# 7 Events for IoT

One of the requirements for IoT-aware business processes defined in the IoT-A Project, is the representable using of event types. Events in business processes are used to address the high flexibility in IoT-processes [10]. An event represents something that occurs during the process. They can either be triggered by something or have an effect on the outcome.

Events can be useful in different scenarios within an IoT-aware business process. However, insufficient research is done into the use of events within IoT. According to Chiu and Wang [18] the main focus of other researchers has been on extending Pool, Activity and Resource. However, many IoT processes could be better modeled using events.

The extended BPMN 2.0 already contains many different event elements for standard business processes, shown in Appendix C. These events can be divided into three main types, Start, Intermediate, and End event. For IoT-aware business processes, this distinction is still relevant. Furthermore, event types can help to address several of the requirements defined for IoT-aware BPs in Chapter 3 of this report. This could be events for mobility, such as a location event, but also the availability of a thing can be modelled using events. Lastly, event types can also be used to express real-time constrains and can be used in the case that errors occur.

Both the Temporal and Error events are not unique for IoT, but therefore not less relevant. Some handlings or communications between physical objects must happen at a certain moment in time, for which timer events can be used, **Figure 8**. The error or failure events are also available in BPMN 2.0 and can be used when something goes wrong during a process,

Figure **9**. In case of an IoT-aware process, this can be failures of IoT devices, such as death batteries or wrongful data sharing.

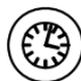

**Figure 8 Timer event**

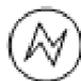

**Figure 9 Error event**

In an IoT-aware process it can occur that communication between physical objects is only allowed when the objects are in a certain state. Furthermore, sometimes a certain handling can only be done when the object is in a state. To ensure that this handling is only executed when the object is in that state an event can be used. In the current literature, a state event for IoT cannot be found.

The location event is used when a certain action can only occur, when an object is at a certain place or within the range of another object. According to Meyer et al.[10] the new element for location event solves the problem of the mobile devices and process in IoT-aware processes. Both Chiu and Wang[18] and Kozel19] introduce graphical notations for location events.

The location event defined by Chiu and Wang is shown in Figure 10. This event is an open circle with a pin in it and can be used as start or intermediate event. It can describe situations in which an entity is at a certain position or is actually moving [18].

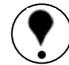

Figure 10 Location event

Kozel [19] introduced additional location based events shown in Figure 11. These event types can be used to express mobility in business processes. The first row present position achieved events. These events are triggered whenever the mobile participant reaches the requested place. The position update event is shown in the second row. When a position change happens which is significant for the process, this event is used. Lastly, the conditional positional event is shown. This is a more general event and can be used when a mobile participant is within a certain range of the location [19].

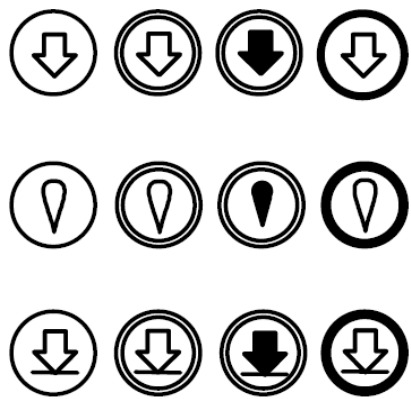

Figure 11 Location based events [19]

Chiu and Wang introduced other events also relevant for IoT-aware business processes, based on the eight requirements of Meyer et al [10]. The Conditional start event is extended so that it can also respond to triggers from IoT Devices. Furthermore, it is no longer only a Start event, but also an intermediate event. Which means that it can take place after the process already started and affects the flow of the process [4]. An extension of the message event is created so that interaction is possible between two separate pools. Lastly, the current Error event is extended, such that after an error occurs an alternative process is started.

The introduction of the event element for IoT will help to design IoT-aware business processes and handles issues as scalability and mobility [18].

# 8 Data

Data plays an important role in IoT-aware business processes. The sensing activity retrieves data from the Physical Entity; this data should be processed and can lead to an actuation activity. Sometimes data only needs to be stored temporarily and must be processed immediately. In other cases, the data is stored beyond the time of a process.

It is important that the data that triggers certain actions and can influence the sequel of the business process is of a good quality. For example, when a patient is controlled from a remote location and his blood pressure drops an alarm goes off. This has a big influence as it actuates the ambulance post and request for the ambulance to take action. When this happens, health workers must be certain that this alarm is based on correct information. Hence, the quality and veracity of information should be taken into consideration for IoT-aware business processes.

Lastly, data should not be available to everyone and it might be useful to show this in a business process. The access privileges, that allow people to change or view data, are managed on a lower level than the BPs. However, it can be relevant to show in a Business Process that certain data is private. For example, patient data in a hospital is considered to be private data, it is important to control if a person has access to it, before the data can be viewed. Research is done on data security and privacy for an IoT setting in the healthcare area. The outcomes might be useful for general use of data security in IoT-aware processes.

## 8.1 Real world data

The data that is part of an IoT-aware process can be either of the type real-world data or of normal data. Normal data here is defined as data from the digital world, which is part of the business process. When the data is gathered from the Physical Entity this data is derived from the real world. BPMN provides two objects to store information, a data object, and a data storage. The data object only stores information while a process is running. The data that is stored in a data storage lasts beyond the lifetime of the process and can be updated by internal and external participants [10].

According to Meyer et al. [10] it is important to make a distinction between real-world data and digital data. It is already possible in the XML format to add information to the data storage, but this is not sufficient, as it does not present a graphical element. Therefore, it is added in the metamodel that a data object or storage can be either of the type real world or not. They also created graphical elements and an adapted XML scheme to show that the stored data is of the type real world, shown in Figure 12 and Figure 13. In the extended versions it is also possible to add properties to is such as the source of the data or the delay time [10].

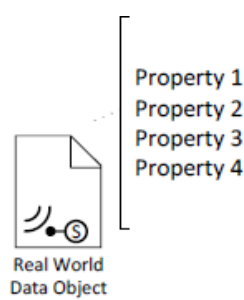
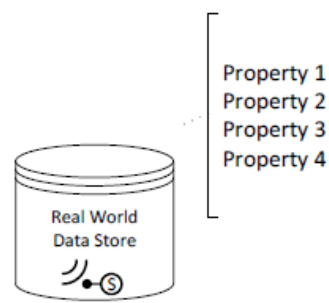

**Figure 12 Extended Real World Data Object**    **Figure 13 Extended Real World Data Storage**

As mentioned before the meta model defined by Meyer et al [10] is too complicated and the new elements to store Real World Data is an important reason. Is it even of importance to a business process to see where the information is going? Chen and Wang [3] argue that data flows are only known by 30% of the BPMN users. Furthermore, it takes a long time to fully understand new BPMN elements. Considering both arguments, extending the BPMN with two elements for real world data is unnecessary. In case that the information flows should be represented in the business process the current elements of BPMN 2.0 are sufficient [3].

Petrasch and Hentschke [20] agreed that data was an important aspect and they created an additional element named cloud application. The cloud application element is added to either a pool or lane, as shown in Figure 14. The cloud application has its own process and operates as a kind of buffer before the data is stored. This element is especially useful when the IoT-aware process also includes cloud computing[20].

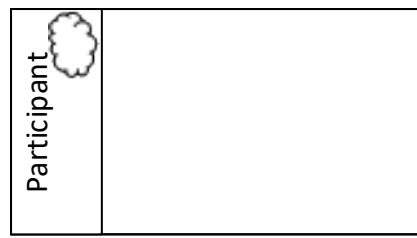

**Figure 14 Cloud Application**

## 8.2 Data quality

In IoT-aware processes, the quality of data is crucial. Businesses have to trust in the data that is sensed from the Physical Entity, as well on the data sent by the Physical Entity after an actuation. This data transport happens directly between sensors and actuators. No human involvement or additional control is executed to see if the data is valid or complete. In case the sensor does not work properly or the base level for actuation is set wrong, problems can occur in the remainder of the process. Until now, business processes did not need to include the quality of data as this was considered as a given fact. Nowadays, with IoT involved, there are more uncertainties. Integrating it into the business model notation makes it possible to deal with inaccuracies beforehand and certain measures can be taken when they occur.

The quality of data is measured at two levels, the Quality of Information (QoI) which characterized information from the sensor to determine the reliability of the Quality of Actuation (QoA) which describes the actuation constraints on how well requests are executed. Combined together they form the IoT Quality Metric (IQM), which measures the quality on four different aspects, traceability, reliability, spatial and temporal. For each of the values a ratio can be calculated.

Each of these parameters is treated separately as a vector, so for example the temporal vector. This vector again is composed of separate elements, such as StartTimeActuation and ExecutionTimeActuation. Each aspect creates its own vector, which is built of specific and separate elements for that aspect. The individual components of this vector are weighted with a separate vector, which is called the temporal weight vector $\vec{g}_{tem}$, in the case of temporal quality. This results in the overall vector of temporal quality.

To calculate the overall IQM ratio the ratio of each quality is divided by the IQM weight factor. The higher the IQM ratio, the better the provided information is. The calculated IQM ratio can also be represented in the business process model. Both a XML schema and a graphical element are created. The graphical element, Figure 15, shows the IQM ratio for a specific Sensing Activity, which represent the quality of the data generated by the specific task [21].

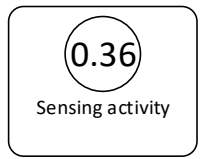

**Figure 15 IoT Quality Metric**

## 8.3 Data security

The increasing usage of IoT within businesses also means that the amount of information that is generated, processed and stored increases. Protection of privacy and important information is of great importance in all areas of business. The current available BPMN notation lacks to present any security concepts. Therefore, an extension is made to include security requirements within BPMN. The research found on IoT and security requirements is focused on the healthcare area. This counts also for the research done by Sang & Zhou [22], which is discussed below.

**Table 3 Security requirements**

| Element | Explanation |
|---|---|
| Authentication | Check if the person is who he says he is, before granting access. |
| Access Control | Access to a place is selective for a group of authenticated users. |
| Authorization | An authenticated person with access to resources is allowed to take further actions |
| Harm Protection | Scans the process to protect the system from malignant attacks. |
| Encrypted Message | The message that is sent is encrypted or signed. |
| Non-Repudiation | There is an agreement on interaction between two roles |
| Secure Communication | Two roles are connected under certain security protections. |

There are three types of security elements, Security Task, Security Event, and Security Indicator. The paper focuses on three main security goals which are Confidentiality, Integrity and Availability. The goals must be reached by fulfilling seven different security requirements, shown in Table 3. These security requirements are presented as a BPMN event with a meaningful icon in it, for example a key for secure communication. The Security Activity is created to show the activities related to security.

The security goals are a new concept within BPMN, which means that a new graphical element is created for it. Figure 16 shows the element. The letter defines the security goal and the amount of stars show the level of security that is required. When a new security goal is necessary for a certain business process it is easy to adapt the element that is created here [22].

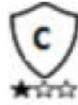

Figure 16 Security Goals

The BPMN extensions for security presented here are derived from the healthcare domain. Within healthcare, protecting the privacy of the clients is of great importance. Thus, it is no surprise that the research focuses on this domain. However, the presented extensions are also of great use for other areas, because every business has to cope with these security measures. Therefore, these extensions can be used in a more general way for IoT-aware processes.

# 9  Location and Mobility

Location is one of the distinguishing characteristics of "things", among other physical characteristics that need to be considered. These characteristics are the most important drivers of extensions to the modeling languages for IOT-aware business processes. Prior – middleware-oriented – business process management systems and their associated modeling languages were focused on execution of administrative processes in which locality and other physical properties were considered as one of many data parameters or in the context of user interaction.

An IoT application is particularly advantageous due to its high level of connectivity, also called its '6A' Connectivity. '6A' Stands for Anyplace, Anything, Anyone, Any Path, Any service and Anytime [23]. Therefore, location (as geographical position at a certain moment in time) and mobility (as change of geographical position) are an important aspect in IoT-aware business. Both an IoT Device and Physical Entity can be mobile. Mobility in an IoT-aware business process influences the behavior of this process. It can be that the location of the participant changes and communication is no longer possible or that a certain decision or action depends on the location of this participant [10].

An IoT network can consist of many IoT devices. These devices can be seen as either dynamic or static nodes. Due to the high connectivity of an IoT application, these nodes should be able to communicate with each other.

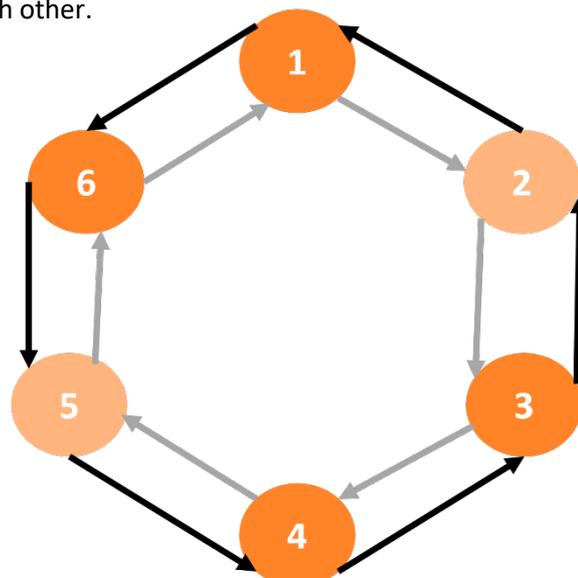

**Figure 17 Mobile IoT Network**

The dynamic nodes 2 and 5 can move along the network, show in Figure 17. They encounter the static nodes and are able to communicate with them for a certain amount of time. The time depends on the available range and speed [23].

The network mentioned above is also known as an opportunistic network. Whenever a node can communicate and share information with another node, it takes this opportunity. This type of networking is especially useful in a scenario's where no end-to-end paths are defined, such as smart cities [23]. To reach the full potential of such an opportunistic network four properties are crucial, namely dynamicity, context-awareness, co-location and transience [24]. For mobility, context-awareness and co-location are important. Transience is discussed in the next section, as this represents time requirements. Context-awareness, means that any information on the entity should be considered, such as location and physical condition. This can be visualized in BPMN with collapsed lanes or text annotations. Co-location, means that different entities can share the same resources at the same location [24]. There is no possibility yet to model this in BPMN.

The influence of mobility discussed above is mainly on the aspect of information sharing, when devices are within a certain range of each other. However, mobility can also be important for the physical handling of things. An example of a location dependent physical handling is for example a parcel delivery. The parcel should only be taken out of the vehicle when the right location is reached. The dynamic states of the IoT devices and physical entities, such as their location and movement direction can influence the process operation [2]. In a mobile process, the process depends on the presence of a certain participant. The current notation BPMN offers for mobility is not sufficient to handle the mobility of all different elements. Although business processes do not support mobility yet, a text annotation could be used to show an element is mobile[10]. However, any elements can be mobile and creating an annotation for each of those can look unclear. For both transparency and agility, the context of mobility needs to be considered in the modelling notation [2].

Meyer et al. [10] created two elements to solve the mobility problem for business processes. The separated arrow shows that an element is mobile while the location marker shows that the artefact is location based (**Error! Reference source not found.**). The mobility element can be added to either a process or a participant. In case of a mobile process, the arrow is shown in the pool and possibly in some of the lanes that the pool contains. In case of a mobile participant, the element is added either to the lane of the participant or to the collapsed pool of the Physical Entity. An activity can either be mobile or location-based, the element is then added to the activity block[10].

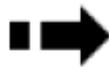  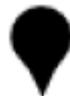

Figure 18 Element for Mobility    Figure 19 Element for Location-Based thing

In Chapter 6 the location based event introduced by Kozel [19] were presented. These additional elements provide a solution to model mobile requirements for positions and places in BPs. Another way to address the mobility issues in business processes is to model mobility as an error or exception. In the case the thing cannot find the right location, an error can be modeled. This means that the domain expert does not have to keep track anymore, but the process model can do it independently. It is important to consider what kind of exception should be modeled. In the case that an entity or device is not at a certain location an error or exception can be explicitly modelled with Java and implemented in the BPMN mechanism[3].

Mobility plays an important role within IoT, as the number of devices and things with GPS massively grow[19]. The mobile participants must be at a certain predefined location or within a certain radius of each other, to perform their activities. This can be either for data sharing between devices or physical handling of the entities. The above-mentioned extensions do provide a solution, to visualize that a certain participant is mobile or that an event is triggered when spatial requirements are fulfilled. However, it does not provide a transparent nor elegant solution that two participants need to be within a certain distance for each other. Nor does it show the locations where certain objects must be before the process can continue. Therefore, there is still a possibility to improve the integration of mobile participants in business processes to make it more transparent and agile.

# 10 Timing

The Internet of Things can be applied in many different application domains, as was discussed in Section 2 of this report. It is often the case that real-time data is used to reach the full potential of IoT in those applications. For example, when delivering a parcel, an Estimated Time of Arrival is communicated with the customer. Real-time data can be used, such that the driver can take the most optimal routes with the least traffic and deliver the parcel at the right time. Using IoT successfully means that the best possible feedback from the system is necessary. Decisions should not be based on historical data, but preferably on real-time data and real-time monitoring [25].

Temporal requirements and spatio-temporal properties are both important concepts for cyber physical systems and are therefore related to IoT networks. In some physical processes, timing is crucial and actions must be taken at the exact right moment in time. Within a process, activities are closely linked and can be time-related. This can include activity duration or time constraints. For a consistent process state, these temporal requirements are crucial in the design of a process[26]. The spatio-temporal properties combine the information from both the location and the time and according to this the state of a physical object can change [27].

An important contribution to temporal modeling is Allen's Interval Algebra. Allen's work describe relations between two activities with respect to time. In total 13 possible temporal relations are described and can be used in modelling time related business processes [28]. A visualization of the relations is presented in Figure 20.

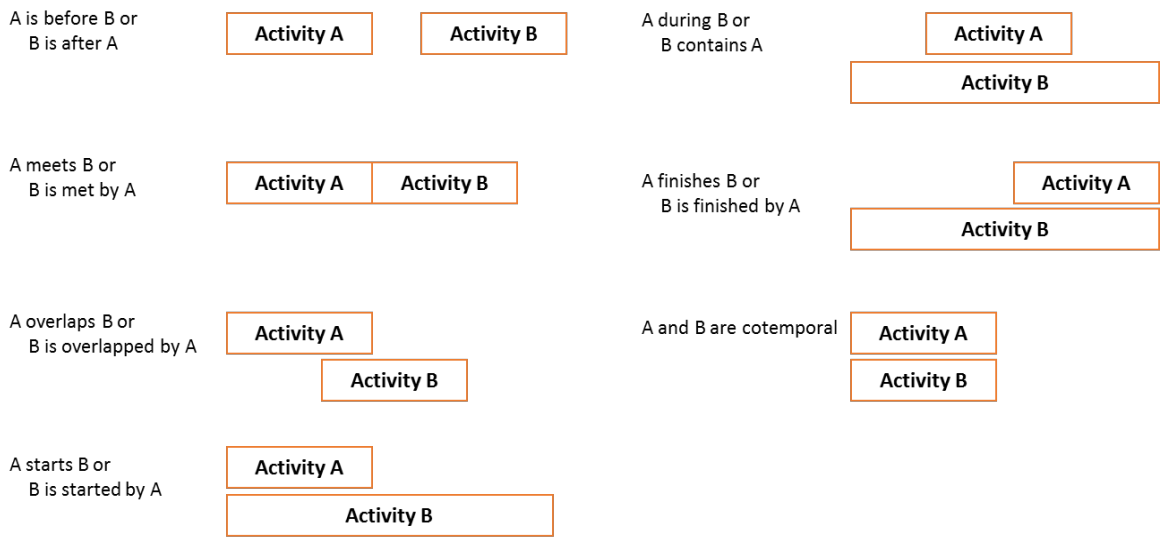

**Figure 20 Visualisation of Allen's intervals**

However, the intervals are mainly a theoretical description of the possibilities and there is no direct implementation possible in any of the standard business modelling languages. Therefore, Kumar [28] introduced a temporal model for business processes that integrates the temporal patterns in BPMN. The only thing necessary is the standard BPMN including Tasks, gateways and arcs and additional Task Durations and Inter-Task constraints. The Task Duration can be determined for each specific task and contains a minimum and/or maximum time of duration. The graphical notation of the Task Duration is shown in Figure 21.

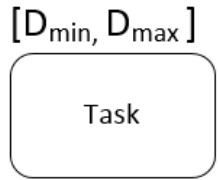

Figure 21 BPMN Extension for Task Duration

The inter-task constraint is more elaborate and variations for this constraint exist. The inter-task defines the minimum and/or maximum duration in between tasks. However, the moment the time measurement start might variate between the start and the finish of a task, which allows for variations of the inter-task constraint. The inter-task constrain can determine that task B should start within 2 hours after task A was started. Another possibility is that Task B should finish at least 100 minutes after task A started. A dashed line and an annotation of the maximum and/or minimum duration represent the inter-task constraint. When the start of a task is the beginning of the time measurement, the arrow points at the front of the rectangle, for the end of a task the arrow points at the back of the rectangle. The various possibilities are visualized in Figure 22.

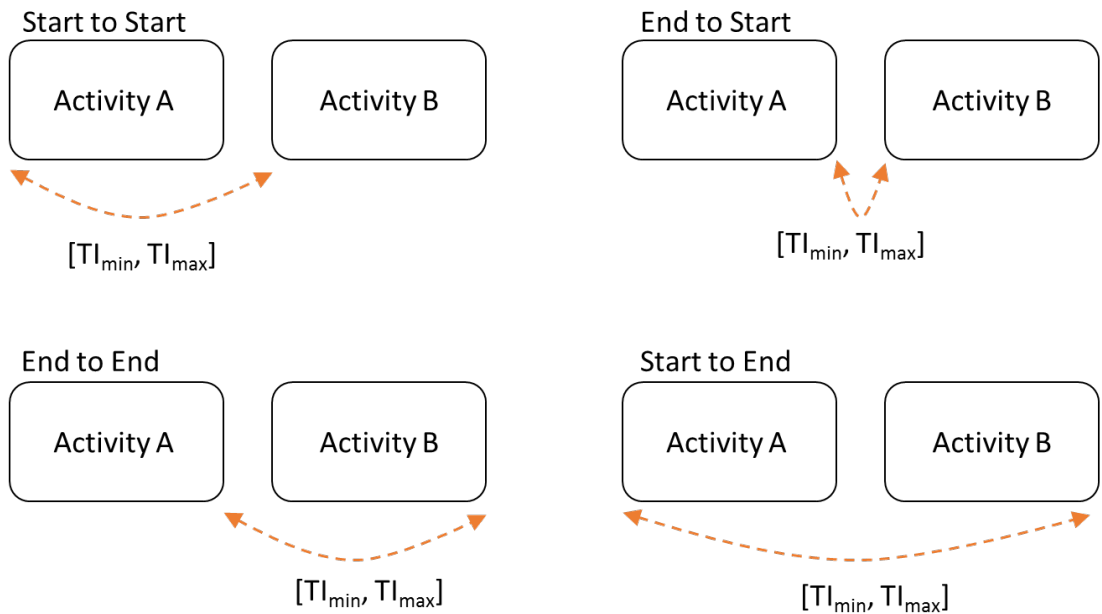

Figure 22 BPMN Extension for Interarrival-Task Constraints

It is already possible to model time using the standard BPMN 2.0. BPMN 2.0 includes a timer event that can set either a specific time, 9 am, or a cycle-time, every 2 hours. The timer can be used as either a start event to trigger the process or an intermediate event that delays the process [4]. The extension of Kumar [28] also enables the expression of the duration of a specific task as well as the minimum and maximum time that should be in between tasks. However, BPMN does not allow for a visualization where two participants should be available at the same time. Neither, is there a concept that presents relative time, between two participants. To fulfill the requirements for an IoT network, another extension to BPMN is necessary.

# 11 Related research

Besides IoT, there are similar approaches that connect the physical world with the digital and which focus on sensor networks. This section focuses on research that is done on related topics.

## 11.1 Web of Things

The Web of Things (WoT) provides a different view on IoT. In the Web of Things, the smart devices are integrated into open web services. The main difference is that the WoT enables the IoT devices to speak the same language, so that they can communicate using the web, instead of giving each device an IP address to connect them to the Internet [29].

The WoT initiative uses the Representational State Transfer (REST) to connect the web services to the physical world. REST models sensors or RFIDs that are important enough for a business process as a resource. The resources are accessed using the uniform interface HTTP. An effort is done to integrate the WoT interactions and elements into the BPMN, which lead to the creation of both graphical and machine-readable elements [30].

## 11.2 Cyber Physical Systems

Cyber Physical Systems (CPS) are the integration of digital and physical processes. Both processes can influence each other's computations [31]. The research done on the integration of CPS in BPMN also shows the creation of the actuating and sensing task. However a main difference is that each business process consists of three lanes that show the physical process, a controller and the cyber process [27]. Furthermore, the Performability-enabled BPMN (PyBPMN) is created. This new notation allows to integrate important CPS properties such as workload and reliability, into the BPMN notation[31]. Lastly, temporal and physical properties are important to ensure that the CPS can perform accurately. For that reason BPMN4CPS is created, which includes these properties into BPMN using graphical elements[27].

## 11.3 uBPMN

Ubiquitous business processes have the capability to use their environment to extract data from it and can change this environment with a minimum human intervention. The ubiquitous business process is location-independent, which means that a ubiquitous process is not attached to one specific location [32]. However, the specific location can be taken into account while executing the business process, as the data used from the environment is different for each location. This is fairly similar to the approach of IoT in terms of mobility and the use of location. The main difference between Ubiquitous computing and IoT is that the first focuses on having computational capability in many objects, while the latter connects objects to the internet [33].

## 11.4 Event Stream Processing Units

IoT produces a large amount of data. Businesses can use this data to improve their processes when they integrate the data from event stream into their processes. During event streams, new events occur over time and the consumer of the event is nonexistent or not known.

Single events are already well known in BPMN, but to deal with the real-world data it is necessary to extend the BPMN elements. Therefore, Event Processing Stream Tasks (ESPT) are introduced to serve as Stream Processing Units and Event Stream Specification (ESS). An ESPT needs an ESS as input when the control flow arrives. The added elements allow the process to have triggered completion, which is required in event processing [34].

## 11.5 Wireless Sensor Networks

Wireless Sensor Networks (WSNs) are IT systems, but have certain properties that make it different. For example, the operations executed by WSNs are more restricted than multipurpose IT systems. WSN is a part of the IoT concept and focuses on the sensors that connect the digital world with the physical world. WSNs are literally the eyes and ears of the IoT.

The *make*Sense project is a European project that created a bridge between the technical part to program sensors and the useful design of processes. WSN business processes can be modeled using the BPMN elements that are currently available. However it is more convenient to adapt the BPMN to cover the specific properties of WSNs in business processes [35].

## 11.6 Benefits and challenges framework

The paper of Janiesch et al. introduces the mutual benefits and challenges for IoT and Business Process management. These challenges occur on both sides. IoT still needs to deal with many technical complications such as computational limitations and connectivity. BPM normally handles and analyzes pre-defined processes. However, when combined with IoT it needs to be able to handle agile process and synchronous actions. This paper opens many different research possibilities in the field of IoT and BPs and only a part of these possibilities are fulfilled yet by the research shown in this state of the art report.

To model ubiquitous business processes an extension to BPMN is created called uBPMN. The extensions are both machine-readable and graphical. The amount of tasks is expanded and distinguishes what type of data is collected and how it can be collected [32].

# 12 Conclusion

The IoT-A project introduced the first IoT-aware business process modelling (IAPM). The IAPM proposes seven new elements that extend the possibilities of BPMN 2.0 for IoT-ware business processes. However, the IAPM is not thoroughly evaluated aat the time of writing this summary and it might not fulfill all the needs to model IoT-aware business process. We need additional research on the evaluating the application of IAPM by to different domains and needs.

Other research proposes alternations to the new elements, such as including a resource role for activity and sensing task. Some created additional elements or decided that some of the elements of the IoT-A project were unnecessary. For example, Appel et al. [34] chose to model IoT-aware processes using event streams in order to use the value of the real-world data that was collected.

This report reviews the state of the art on modelling IoT-aware business processes in BPMN. The extensions to BPMN are elaborated and show a solution for the main issues that were of concern to IoT-aware business processes. However the new modelling elements have not been evaluated thoroughly yet and the question remains if they are sufficient when modelling the processes.

Furthermore, in the current research the thing is treated as a black box. The assumption is made that a physical entity cannot have its own process and therefore the representation as a black box is considered sufficient. However, this assumption is questionable and even if it would be true, a physical entity can have its own lifecycle. The lifecycle contains states in which the thing can be. In the case of mobile things, a state can also be a location. The states in which a thing stays can change due to influences from the process, such as actuating activities. Representing the states can be important for both clarification of the business process and the accuracy of the remainder of the process. The current modelling extensions do not enable to make distinctions between different states of the things. In future research it would be interesting to see if this is necessary and how this can be modelled in a business process.

Lastly, there is not much work on spatial and temporal synchronization of things. The use cases, analysis and research shows that special and temporal coordination is necessary for IoT-aware business processes, no extension for BPMN is found in this field at the time of writing to the best of our knowledge. This is a promising area of future work.

# 14 Appendix A: BPMN Elements

**Table 4 BPMN Elements [4]**

| Element | Name | Explanation |
|---|---|---|
| 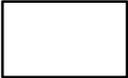 | Activity | A task that is performed in a process |
| 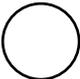 | Event | An occurrence during the process |
| 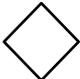 | Gateway | Controls the split or join of a sequence flow. This can mean that both flows must be executed or only one. |
| 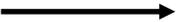 | Sequence Flow | Shows the order of activities that must be performed in the process |
| 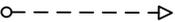 | Message Flow | Shows message exchange between participants |
| 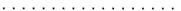 | Association | Link data objects with BPMN elements |
| 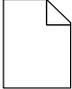 | Data Objects | Information storage of a single activity |
|  | Message | Communication between participants |
| 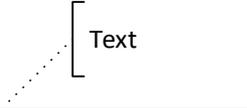 | Text Annotation | Provides extra information in the form of a text |
| 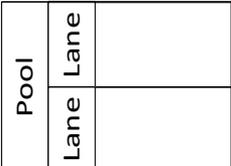 | Pool/Lane | A pool represents the collaboration of participants. Each lane represents a participant and includes a part of the process. |

# 15 Appendix B: IoT concepts for Business Process Models

| Element | Name | Explanation |
| --- | --- | --- |

**Table 5 New Concepts for IoT-aware business process modelling**

| | | |
|---|---|---|
| 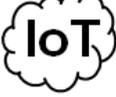 | IoT Device | Element that specifies that a participant is an IoT Device |
| 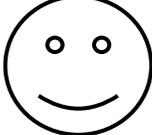 | Human Involvement | Element that specifies the human involvement in a process |
| 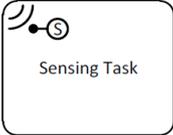 | Sensing Activity | A sensing task that is performed in a process |
| 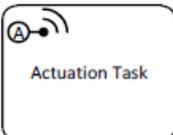 | Actuation Activity | Shows the order of activities that must be performed in the process |
| 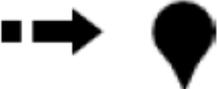 | Mobility/Location-based | Element that identifies a participant or activity as mobile or location-based |
| 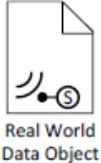 | Real world data object/store | Stores real-time information of an activity |
| 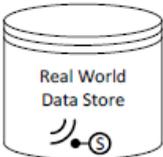 | Cloud Application | Processes data before it is stored |
| 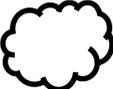 | IoT Quality Metric | Shows the reliability of data collected by an activity |
| 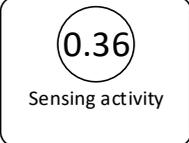 | Security Goal | Shows the security requirements for a certain part of the process |
| 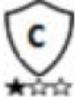 | Location Event | Intermediate event to describe the position of an entity |

# 16 Appendix C: BPMN 2.0 Events

| Events | Start | | | Intermediate | | | | End |
|---|---|---|---|---|---|---|---|---|
| | Standard | Event Sub-Process Interrupting | Event Sub-Process Non-Interrupting | Catching | Boundary Interrupting | Boundary Non-Interrupting | Throwing | Standard |
| **None**: Untyped events, indicate start point, state changes or final states. | ○ | | | | | | ○ | ○ |
| **Message**: Receiving and sending messages. | ✉ | ✉ | ✉ | ✉ | ✉ | ✉ | ✉ | ✉ |
| **Timer**: Cyclic timer events, points in time, time spans or timeouts. | ⏲ | ⏲ | ⏲ | ⏲ | ⏲ | ⏲ | | |
| **Escalation**: Escalating to an higher level of responsibility. | | ⋀ | ⋀ | | ⋀ | ⋀ | ⋀ | ⋀ |
| **Conditional**: Reacting to changed business conditions or integrating business rules. | ▤ | ▤ | ▤ | ▤ | ▤ | ▤ | | |
| **Link**: Off-page connectors. Two corresponding link events equal a sequence flow. | | | | ➡ | | | ➡ | |
| **Error**: Catching or throwing named errors. | | ⚡ | | | ⚡ | | | ⚡ |
| **Cancel**: Reacting to cancelled transactions or triggering cancellation. | | | | | ⊗ | | | ⊗ |
| **Compensation**: Handling or triggering compensation. | | ⏪ | | | ⏪ | | ⏪ | ⏪ |
| **Signal**: Signalling across different processes. A signal thrown can be caught multiple times. | △ | △ | △ | △ | △ | △ | ▲ | ▲ |
| **Multiple**: Catching one out of a set of events. Throwing all events defined | ⬠ | ⬠ | ⬠ | ⬠ | ⬠ | ⬠ | ⬟ | ⬟ |
| **Parallel Multiple**: Catching all out of a set of parallel events. | ⊕ | ⊕ | ⊕ | ⊕ | ⊕ | ⊕ | | |
| **Terminate**: Triggering the immediate termination of a process. | | | | | | | | ⬤ |

**Figure 23 BPMN events** [18]